\documentclass[fleqn,twoside]{article}
\usepackage{espcrc2}
\usepackage{amsmath}
\usepackage{graphicx}

\title{Nonstandard Cutoff Effects in the Nonlinear Sigma 
 Model\thanks{Presented by B.S.}}
\author{
 M.~Hasenbusch\address{NIC/DESY Zeuthen, Platanenallee 6, D-15738, Germany}, 
 P.~Hasenfratz\address{Institute for Theoretical Physics,
    University of Bern, Sidlerstrasse 5, CH-3012 Bern, Switzerland}, 
 F.~Niedermayer{${}^{\rm b}$}, 
 B.~Seefeld{${}^{\rm b}$}, 
 U.~Wolff\address{Humboldt Universit\"at zu Berlin, Institut f\"ur
    Physik, Invalidenstrasse 110, D-10115 Berlin, Germany}}

\begin{document}

\begin{abstract}
High precision measurements of the renormalized zero-momentum 
4-point coupling $g_R$ and of the L\"uscher-Weisz-Wolff running
coupling $\bar{g}(L) = L m(L)$ performed with two different
lattice actions in the non-perturbative region confirm the
earlier observations, that the cutoff effects look linear, in contrast
to perturbative considerations. The use of different actions allows one
to make a more reliable estimate on the continuum limit. 
The measurements were done for infinite volume correlation length up
to 350.
\end{abstract}

\maketitle

In numerical simulations with lattice regularization the physical results 
are obtained after extrapolating to zero lattice spacing.
Usually (and especially in 4d) the range of $a$ values accessible in
the simulations is quite restricted, therefore the extrapolation 
$a\to 0$ is highly non-trivial and the knowledge of the functional form 
of the artifacts is essential. 
As shown by Symanzik \cite{Symanzik}, in bosonic theories in every 
order of perturbation theory (PT) the leading lattice artifacts decrease
like O($a^2$) (apart from $\log a$ factors).
It is generally assumed that this behaviour holds beyond PT, and the
extrapolations in numerical simulations are done according to this
assumption. 

In \cite{Hasenfratz:2001sa} the renormalized zero-momentum 4-point 
coupling was measured in the O(3) non-linear sigma model to high
precision. Unexpectedly, the cutoff effects were nearly linear in $a$
and were described well by a form $c_0 + c_1 a + c_2 a \log a$,
in contrast to the standard belief.
(Note that already in \cite{Balog:1999ww} there was some numerical
evidence that a linear Ansatz describes the data better, but the
errors were significantly larger.)

Here we extend the results of \cite{Hasenfratz:2001sa} to another
quantity, the L\"uscher-Weisz-Wolff running coupling $\bar{g}(L)$
\cite{Luscher:1991wu}. We also measure these quantities using
an alternative (ad hoc) lattice action containing a diagonal 
interaction term.
For this action the coefficient of the nearest neighbour (on-axis)
term is $\beta_1$ while the coefficient of the diagonal term is $\beta_2$.
They were chosen to be equal, $\beta_1=\beta_2=\beta/3$.
(In this notation for the standard action $\beta_1=\beta$,
$\beta_2=0$.)

The simulations were done with both actions at the same values of
the correlation length and lattice size (i.e. same $a/L$ and $\xi(L)$).
The assumption of universality implies that both actions lead to
the same continuum limit for physical quantities. 
This strongly restricts the fits to the data.

The quantities measured  are:

1) The renormalized 4-point coupling at finite volume
\begin{equation}
g_R(z)=\left(\frac{L}{\xi_2(L)}\right) 
\left( \frac{5}{3} 
-\frac{\langle \left( \mathbf{M}^2 \right)^2 \rangle}{
\langle \mathbf{M}^2 \rangle^2} \right) \,,
\end{equation}
defined on an $L\times L$ periodic lattice.
Here $\mathbf{M}=\sum_x \mathbf{S}$ is the total magnetization
while $\xi_2(L)$ is the second-moment correlation length 
(cf. e.g. \cite{Balog:1999ww}). 
The physical size, $z=L/\xi_2(L)$ is kept fixed.

2) The running of the LWW coupling \cite{Luscher:1991wu}. 
The coupling $\bar{g}(L)=m(L) L$ is defined on an $L\times \infty$
strip, where $m(L)$ is the inverse correlation length in the
zero-momentum spin-spin correlation function.
For a given number $L/a$ of lattice sites in the spatial direction
one tunes the value of $\beta$ to get a fixed value 
$\bar{g}(L)=u_0$. In the next step one measures 
$\bar{g}(2L)$ at the same value of $\beta$, giving
$\bar{g}(2L)=\Sigma(2,u_0,a/L)$ \cite{Luscher:1991wu}.
The physical (regularization independent) ``step scaling function''
is given by the limit $\lim_{a\to 0} \Sigma(2,u_0,a/L)=\sigma(2,u_0)$.
Due to an improved estimator by Hasenbusch \cite{Hasenbusch:1995rv}
(suited for the strip geometry) we could significantly improve the error.

\begin{figure}[ht]
\includegraphics[scale=0.40]{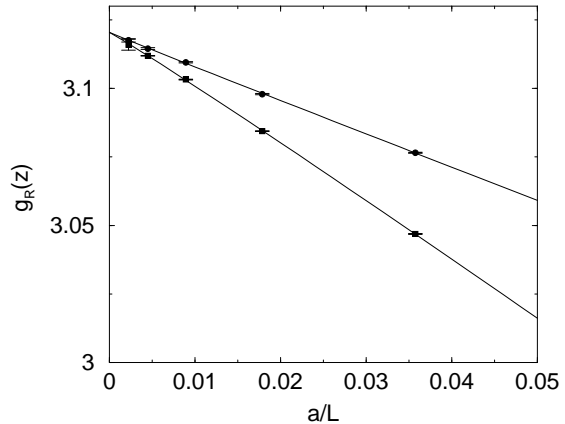}
\vspace{-0.5cm}
\caption{{}$g_R(z)$ at $z=2.32$. The data for the larger
artifacts correspond to the modified action. The fit shown has $a$
and $a\log a$ terms.}
\label{fig:gR}
\end{figure}

\begin{figure}[ht]
\includegraphics[scale=0.40]{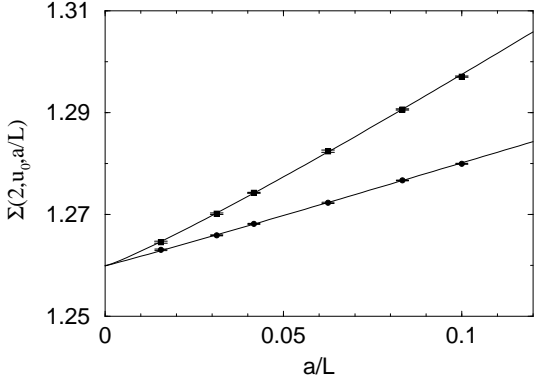}
\vspace{-0.9cm}
\caption{{}The step scaling function $\Sigma(2,u_0,a/L)$ at $u_0=1.0595$. 
The data for the larger artifacts correspond to the modified action. 
The fit contains $a$ and $a\log a$ terms.}
\label{fig:lww}
\end{figure}

The values of $g_R(z_0)$ at $z_0=2.32$ for different lattices sizes $L/a$
and the two actions are given in Table~\ref{table:gR}. 
These data are shown in Figure~\ref{fig:gR}.
Table~\ref{table:lww} and Figure~\ref{fig:lww} refer to the LWW step 
scaling function $\Sigma(2,u_0,a/L)$ at $u_0=1.0595$.

\begin{table}
\begin{tabular}{rcc}
\hline
$L/a$ & (st)      & (mod) \\
\hline
 10   & 3.0105(1) & 2.9237(1) \\
 28   & 3.0765(2) & 3.0469(1) \\
 56   & 3.0979(2) & 3.0844(1) \\
112   & 3.1095(2) & 3.1032(1) \\
224   & 3.1145(3) & 3.1119(1) \\
448   & 3.1175(6) & 3.1159(2) \\
\hline
\end{tabular}
\caption{{}$g_R(z_0)$ for the two actions at $z_0=2.32$.} 
\label{table:gR}
\end{table}

\begin{table}
\begin{tabular}{rll}
\hline
$L/a$ & ~~~~(st)      & ~~~(mod) \\
\hline
  5   & 1.29379(8)  & 1.33591(12) \\
 10   & 1.27994(9)  & 1.29706(14) \\
 12   & 1.27668(9)  & 1.29060(13) \\
 16   & 1.27228(12) & 1.28242(21) \\
 24   & 1.26817(9)  & 1.27424(10) \\
 32   & 1.26591(9)  & 1.27014(15) \\
 64   & 1.26306(16) & 1.26456(20) \\
\hline
\end{tabular}
\caption{{}$\Sigma(2,u_0,a/L)$ at $u_0=1.0595$.} 
\label{table:lww}
\end{table}

The data with $L/a > 10$ were fitted simultaneously for the two actions
by several forms:

\begin{align}
& {\rm A}:\quad  c_0 + c_1 \frac{a}{L} +
  c_2\frac{a}{L}\log\frac{L}{a}\,, \nonumber \\
& {\rm B}:\quad  c_0 + c_1 \left( \frac{a}{L}\right)^2
           + c_2 \left(\frac{a}{L}\right)^2 \log \frac{L}{a} \,,
                                                      \label{Ansatz} \\
& {\rm C}:\quad  c_0 + c_1 \left( \frac{a}{L}\right)^{c_2} \,. \nonumber 
\end{align}

The coefficients $c_0$, $c_1$, $c_2$ correspond to the standard action
while those for the modified action we denote by 
$c_0'$, $c_1'$, $c_2'$.
Universality implies $c_0'=c_0$, and for the
case C we also took the same power, $c_2'=c_2$.
Table~\ref{table:fits} gives the coefficients and the values of 
$\chi^2/{\rm dof}$ for the fits. (The pure $a^2$ fits have 
$\chi^2/{\rm dof} = 11800/7$ and $1270/7$  hence are not listed.)

\begin{table}[ht]
\hspace{-0.5cm}
\begin{tabular}{ccccc}
\hline
 fit    & $c_0$     & $c_1$  &  $c_2$    & $\chi^2/D$ \\
        &           & $c_1'$ &  $c_2'$   &                     \\
 \hline
A       & 3.12121(7) & $-1.266(6)$ &   ---      & $86/7$  \\
        &            & $-2.075(4)$ &   ---      &         \\
A       & 3.1204(2)  & $-1.17(6)$  & $-0.02(2)$  & $18/5$  \\
        &            & $-2.32(4)$  & $0.08(2)$  &         \\
%B       & 3.1085(6)  & $-25.4(2)$  &   ---      & $11800/7$ \\
%        &            & $-49.9(1)$  &   ---      &         \\
B       & 3.11568(9) & $87(3)$     & $-35(1)$   & $180/5$ \\
        &            & $165(2)$    & $-65.8(6)$ &         \\
C       & 3.1201(2)  & $-1.42(3)$  & $1.043(7)$ & $44/6$  \\
        &            & $-2.36(5)$  & $=c_2$     &         \\
\hline
A       & 1.2591(1)  & $0.213(2)$  &   ---      & $74/7$  \\
        &            & $0.372(2)$  &   ---      &         \\
A       & 1.2599(4)  & $0.22(2)$   & $-0.007(8)$& $5.2/5$  \\
        &            & $0.46(2)$   & $-0.038(8)$&         \\
%B       & 1.2653(1)  & $1.64(2)$   &   ---      & $1270/7$\\
%        &            & $3.85(2)$   &   ---      &         \\
B       & 1.2619(2)  & $-2.9(3)$   & $2.0(1)$   & $11/5$  \\
        &            & $-6.5(3)$   & $4.3(1)$   &         \\
C       & 1.2607(2)  & $0.28(1)$   &$1.14(2)$   & $18/6$  \\
        &            & $0.51(2)$   & $=c_2$     &         \\
\hline
\end{tabular}
\caption{{}Fits of the form \eqref{Ansatz} for $g_R(z_0)$ and
$\sigma(2,u_0)$.}
\label{table:fits}
\end{table}

Interestingly, one does not necessarily need to assume a specific
analytic form of the cutoff dependence. Since we performed the
measurements with the two actions at the same values of $a/L$ and
$\xi(L)$ it is enough to assume that the form of the leading 
cut-off effect is the same, only the prefactors are different. 
A quantity $q$ measured with the two actions is given then by
\begin{align}
& q(a)   = q_0 + f(a) + \ldots \label{fifit} \\
& q'(a)  = q_0 + c f(a) + \ldots \nonumber
\end{align}
where $f(a)$ is an unspecified function. We assume that at the values 
of $a$ considered the non-leading artifacts are negligible.
Measuring the quantity $q$ at different values 
$a_1,a_2,\ldots,a_N$ one has $2N$ data $q(a_i)$, $q'(a_i)$
for the two actions and only $N+2$ unknown variables in the fit,
$q_0$, $c$ and $f(a_i)$. 
Both $g_R(z_0)$ and $\sigma(2,u_0)$ are described well by the
assumptions made in \eqref{fifit}.
In Table~\ref{table:fif} we give the continuum value, $c$
and  $\chi^2/{\rm dof}$ for the two quantities 
measured.\footnote{Obviously, by taking three different actions 
one can resolve the next-to-leading cutoff term as well.}

\begin{table}
\begin{tabular}{cccc}
\hline
                & $q_0$     & $c$     & $\chi^2/D$ \\
\hline
$g_R(z_0)$      & 3.1181(5) & 1.71(1) & $6.7/3$ \\
$\sigma(2,u_0)$ & 1.2614(3) & 1.91(2) & $1.0/3$ \\
\hline
\end{tabular}
\caption{{}Results of the form-independent fits.}
\label{table:fif}
\end{table}

For small $\bar{g}(L)$ the cutoff effects can be studied in PT.
The sign of artifacts at  $\bar{g}(L)={\rm O}(1)$ 
(like in Fig.~\ref{fig:lww}) is opposite to that obtained in PT.
This has been noticed already in \cite{Luscher:1991wu}, where
it has also been pointed out that the large $N$ limit shows
the same effect. Caracciolo et al. \cite{Caracciolo}
studied the large $N$ integrals analytically and found 
$(a/L)^2 \left( \log L/a\right)^{-q}$ artifacts with $q=-1,0,1,2,\ldots$
in the step scaling function. They pointed out that 
the negative powers of $\log L/a$ are not reproduced by PT.

Based on our numerical study in O(3) it can not be ruled out that
for very small $a$ the perturbative prediction ($a^2$ times
powers of $\log a$) takes over. In this case an explanation 
would be needed why this happens at $\xi > 350$ only.

\vspace{0.5cm}
\noindent{Acknowledgements:}

We thank Andrea Pelissetto and Peter Weisz for useful discussions. 

This work has been supported in part by the Schweizerischer
Nationalfonds and the European Community's Human Potential Programme 
under contract HPRN-CT-2000-00145. 
We thank IOGRAM AG, Zurich for computing resources.

\end{document}